\documentclass[journal,onecolumn]{IEEEtran}

\ifCLASSINFOpdf
   \usepackage[pdftex]{graphicx}
\else
   \usepackage[dvips]{graphicx}
\fi
\usepackage[cmex10]{amsmath}

\usepackage{graphics} 
\usepackage{graphicx}
\usepackage{subcaption}
\usepackage{array}
\usepackage{cite}
\usepackage{epsfig} 
\usepackage{amsmath} 
\usepackage{amssymb}  
\usepackage{amsthm}
\usepackage{color}
\usepackage{float}
\usepackage[noabbrev]{cleveref}
\usepackage{algorithmic}
\usepackage{url}
\usepackage{calrsfs}
\usepackage{textcomp}
\usepackage{multirow}
\usepackage{booktabs}
\usepackage{bm} 

\usepackage[dvipsnames]{xcolor}

\DeclareMathAlphabet{\pazocal}{OMS}{zplm}{m}{n}
\DeclareMathAlphabet{\mathpzc}{OT1}{pzc}{m}{it}
\newcolumntype{P}[1]{>{\centering\arraybackslash}p{#1}}

\newcommand\ChangeRT[1]{\noalign{\hrule height #1}}

\usepackage[linesnumbered,ruled,vlined]{algorithm2e}
\SetKwInput{KwInput}{Input}                
\SetKwInput{KwOutput}{Output}              
\SetKwInput{KwData}{Parameters}              
\usepackage[cal = pxtx, scr = dutchcal]{mathalpha}

\begin{document}
%
\title{SAVER: Safe Learning-Based Controller for Real-Time Voltage Regulation}

\author{Yize Chen, Yuanyuan Shi, Daniel Arnold and Sean Peisert

	\thanks{Y. Chen, D. Arnold and S. Peisert are with the Lawrence Berkeley National Laboratory,  emails: \{yizechen, dbarnold, sppeisert\}@lbl.gov. Y. Shi is with the University of California San Diego, email: yyshi@eng.ucsd.edu.} 
}

\maketitle

\begin{abstract}
Fast and safe voltage regulation algorithms can serve as fundamental schemes for achieving a high level of renewable penetration in the modern distribution power grids. Faced with uncertain or even unknown distribution grid models and fast-changing power injections, model-free deep reinforcement learning~(DRL) algorithms have been proposed to find the reactive power injections for inverters while optimizing the voltage profiles. However, such data-driven controllers can not guarantee satisfaction of the hard operational constraints, such as maintaining voltage profiles within a certain range of the nominal value. To this end, we propose SAVER: \underline{SA}fe \underline{V}oltag\underline{E} \underline{R}egulator, which is composed of an RL learner and a specifically designed, computational efficient safety projection layer. SAVER provides a plug-and-play interface for a set of DRL algorithms that guarantees the system voltages to be within safe bounds. Numerical simulations on real-world data validate the performance of the proposed algorithm.

\end{abstract}


\section{Introduction}
\label{sec:intro}
The voltage regulation problem has been investigated for decades, yet the increasing penetration from distributed renewable resources keeps adding new challenges to such foundational control tasks. With the greater fluctuations coming from active power injections (e.g., rooftop solar panels and electric vehicles), along with the high $r/x$ ratios of transmission lines, unacceptable voltage swings may appear in the current distribution grids~\cite{carvalho2008distributed}. While on the other hand,  smart inverters of fast-acting distributed energy resources (DERs) can provide reactive power injections in real-time, which can be systematically designed to optimize the voltage profiles~\cite{photovoltaics2018ieee}. 

Previous efforts on the voltage regulation problem focus on designing either centralized or decentralized controllers with optimality guarantees with exact grid models~\cite{qu2019optimal, farivar2011inverter}. This requires distribution grid system operators to either know the exact topology and line parameters or take extra steps to identify such modeling knowledge~\cite{deka2017structure}. With the increasing availability of grid measurements and sensing data, there is a growing interest in designing model-free, data-driven voltage regulation algorithms such as reinforcement learning~(RL) to achieve real-time decision making~\cite{chen2020input, xu2019optimal}.

The RL training process holds the promise of finding control policies with good performance in terms of minimizing the voltage deviations and regulating costs, while it does not need explicit knowledge of the grid parameters. By aggregating nodal voltage and active power injections as the states, the RL agent is trained by interacting with the power grid environment by using learned reactive power injections as actions. However, compared to the model-based counterparts~\cite{farivar2011inverter}, RL is trained to maximize the accumulated reward, while hard physical constraints such as feasible voltage magnitudes are mostly not guaranteed to be satisfied. Unsafe reactive power injections can cause severe impacts over the grids such as voltage collapse and load shedding~\cite{vandoorn2013voltage, chen2021understanding}.

Recent works seek to promote RL safety by reward function shaping~\cite{huang2021guided}, whereby combining constraint violation information, the RL agent learns to avoid unsafe actions and states as such can lead to a high penalty. However, in the training process agents still have to violate the safety constraint and receive the penalty multiple times before it learns to avoid it. In addition, safety is more like an implicit regularization in such methods, as violations of the safety constraint can lead to high costs, while it does not always guarantee safety during implementation. In \cite{liu2021online, wang2019safe}, the voltage regulation problem is modeled as constrained Markov Decision Process~(CMDP) to constrain the state in safe space probabilistically, yet no formal guarantees can be made regarding the voltage profiles in the training or implementation stage.
\begin{figure}[t]
	\centering
	\includegraphics[width=0.7\linewidth]{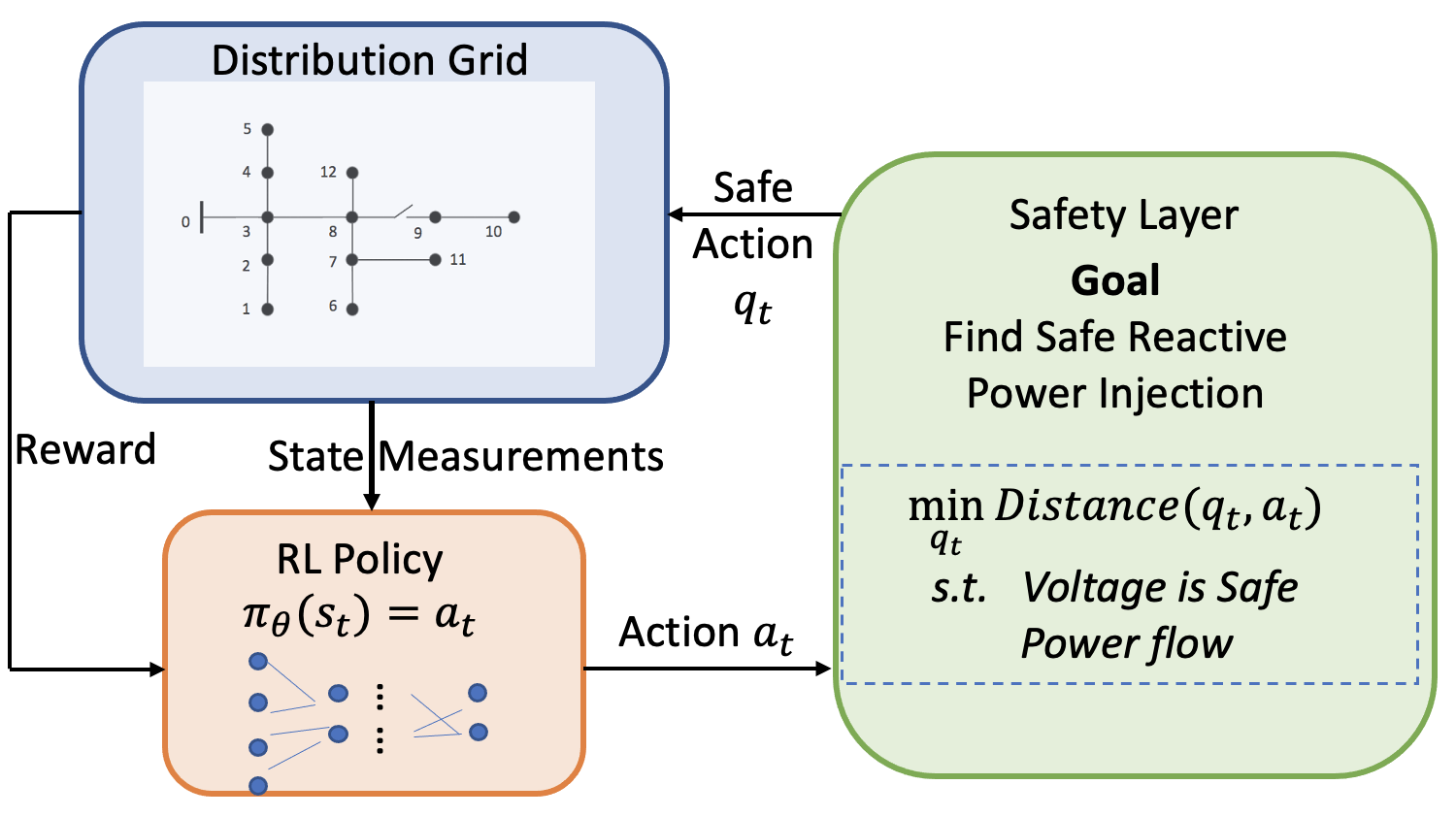}
	\caption{\footnotesize The proposed learning-based control strategy to safely regulate distribution grid voltage.}
	\label{fig:schematic}
\end{figure}

In this paper, we propose SAVER, which is composed of a novel safety layer to overcome such safety concerns regarding RL voltage regulators. By making use of the underlying grid information, we design a projection layer that projects the reactive power injection outputted by the trained RL policy into a safety set of nodal voltage magnitudes. The scheme of the resulting procedure is illustrated in Fig. \ref{fig:schematic}.
The proposed method can fast compute \emph{safe reactive power injections in terms of voltage constraints with guarantees}. Moreover, the safety learning framework can be embedded as a lightweight plug-and-play module for \emph{most if not all} standard reinforcement learning algorithms. 
This work is partly inspired by \cite{dalal2018safe}, where a safe exploration strategy is proposed for reinforcement learning agents in the area of robotic control. We also note that\cite{kou2020safe} proposed a linear projection for safe voltage control policy, yet it only works for the special case where only one single node's voltage hits the safety boundary, while this paper's safety layer works for the general voltage safety set.

The advantages of applying the proposed data-driven voltage regulator are multi-fold. First, it utilizes the empirically proven advantages of
RL algorithms to search the space of neural network parameterized voltage controllers, that can optimize
the voltage derivations and control costs without the exact grid information. Second, it limits the search of the safe policy only to local control actions within the neighborhood of a trained RL policy, adding little computation burden for the resulting controller. Third, the use of such a safety layer encodes the physical constraints, and opens the door for designing a practical controller for inverters.

\section{Preliminaries: Power Flow Models and Problem Formulation}
\label{sec:system}
In this work, we focus on the voltage regulation model for a radial distribution network. We use the graph $\mathcal{G}=(\mathcal{N} \cup\{0\}, E)$ to represent a  radial distribution feeder with $N+1$ buses with bus $0$ denoting the reference bus, and $E$ is the set of connected transmission lines. we denote $\mathbf{p}\in \mathcal{R}^N$ and $\mathbf{q}\in \mathcal{R}^N$ as the nodal active and reactive power injections respectively.

The relationship between the voltage and power injections can be stated as power flow equations. For each line, denote $s_{ik}=p_{ik}+jq_{ik}$ as the complex power flow from bus $i$ to bus $k$, and denote the line impedance as $z_{ik}=r_{ik}+jx_{ik}$. We adopt the DistFlow formulation~\cite{baran1989optimal} as follows
\begin{subequations}\label{eq:power_flow_dist}
\begin{align}
-p_{j} &=p_{i j}-r_{ij} l_{ij} - \sum_{k:(j, k) \in E} p_{j k}, j=1, ...,n \\
-q_{j} &=q_{i j}-x_{ij} l_{ij} - \sum_{k:(j, k) \in E} q_{j k}, j=1, ...,n \\
v_{j} &=v_{i}-2\left(r_{i j} p_{i j}+x_{i j} q_{i j}\right) + (r_{ij}^2 + x_{ij}^2) l_{ij}\,, (i, j) \in E\\
l_{ij} & = \frac{p_{ij}^2 + q_{ij}^2}{v_i}
\end{align}
\end{subequations}
where $l_{ij} = |I_{ij}|^2$, $v_i = |V_i|^2$. Equation~\eqref{eq:power_flow_dist} defines a nonlinear relationship between the active power injection $\mathbf{p}$, reactive power injection $\mathbf{q}$, and the nodal voltage magnitude $\mathbf{v}$.

We consider our control devices are inverter-based DERs (such as renewable generators, battery energy storage), that can change their reactive power output $\mathbf{q}$ in a fast timescale to provide voltage regulation. Denote the set of controllable $q_i, i\in \mathcal{C}$ using the set of nodes $\mathcal{C}$. At each time step, the optimal voltage regulation problem can be then formulated as follows: 
\begin{subequations}\label{eq:opt_voltage_ctrl}
\begin{align}
\min _{\boldsymbol{q}} \quad & \sum_{i \in \mathcal{C}} c_{i}^q\left( q_{i}\right)+ \eta \sum_{i \in \mathcal{N}} c_{i}^v\left(v_i\right)& \label{eq:opt_obj}\\
\text { s.t. } \; 
& \underline{v}_{j} \leq v_{j}(\boldsymbol{p,q}) \leq \bar{v}_{j}, & j \in \mathcal{N}. \label{eq:opt_voltage_limit}
\end{align}
\end{subequations}
The control objective ~\eqref{eq:opt_obj} is to reduce the total control cost for controllable inverters plus the penalty on voltage deviation for all buses, with $\eta$ as a hyperparameter that balances the weights of the two costs. System operators can choose different function forms of $c_{i}^q(\cdot)$ and $c_{i}^v(\cdot)$ to achieve operational goals.
Equation~\eqref{eq:opt_voltage_limit} is the voltage safety constraints that should be ensured at each time step. 

The challenge of directly solving \eqref{eq:opt_voltage_ctrl} lies in the fact that even though we can design a convex cost function (e.g., a quadratic cost over reactive power injection and squared loss of voltage deviation), the underlying power flow \eqref{eq:power_flow_dist} is nonlinear and non-convex, making directly solving the optimization problem involving \eqref{eq:opt_voltage_limit} a hard problem. For the convex relaxation methods based on SOCP or SDP formulations, even though the resulting optimization problem is tractable, it needs the exact information on grid topology and line parameters. In addition, it can take a significant amount of time to solve the optimization problem for large grids, which may not fulfill the goal of achieving fast timescale voltage regulation. These challenges lead to the design need for a computationally efficient model-free controller.

\section{Learning a Voltage Regulator}
\label{sec:RL}
In this section, we describe how we formulate the voltage regulation problem as a RL problem, and illustrate the need for explicitly incorporating safety as a constraint for standard RL algorithm.

RL provides a powerful paradigm for solving \eqref{eq:opt_voltage_ctrl}, in the sense that during the training process, we can train a policy network that maps the states to reactive power injections, to minimize the control objectives defined by~\eqref{eq:opt_obj}. 
First, we define a Markov Decision Process (MDP) of 4-tuple $(\mathcal{S}, \mathcal{A}, \mathcal{P}, r)$ to represent the voltage control model. The states and actions for timestep $t$ can be defined as,
\begin{subequations}
\begin{align}
    \mathbf{s}(t) &:= ((v_i)_{i \in \mathcal{N}}(t), (p_i)_{i \in \mathcal{N}}(t)); \\
    \mathbf{a}(t) &:= ((q_i)_{i \in \mathcal{C}}(t)).
\end{align}
\end{subequations}

Without loss of generality, in this paper, we use $\mathbf{s}(t), \mathbf{a}(t)$ and $\mathbf{s}, \mathbf{a}$  interchangeably. The state transition model $\mathcal{P} : \mathcal{S} \times \mathcal{A} \rightarrow \mathcal{S}$, is determined by the power flow equations~\eqref{eq:power_flow_dist} and external dynamics such as renewable generation and nodal demand; and $r: \mathcal{S} \times \mathcal{A} \rightarrow \mathbb{R}$ is a scalar function that is defined as follows
\begin{align}
    r(\mathbf{s}(t), \mathbf{a}(t)) = ||\mathbf{v}(t)-\mathbf{v}_0||_2^2 + \eta ||\mathbf{q}(t)||_2^2;
\end{align}
where $\mathbf{v}_0=v_0\cdot \mathbf{1}$ and $v_0$ is the feeder head voltage. 

To summarize, given MDP model and an initial policy $\boldsymbol{\pi}_\theta$, the RL for optimal voltage control problem is formulated as
\begin{subequations}
\label{eqn:rl_operation}
\begin{align}
\max_{\theta} \quad & J(\theta):= E_{\mathcal{P}, \pi} \left[\sum_{t=1}^{\infty} \gamma^t r(\mathbf{s}(t), \mathbf{a}(t))\right] \\
\text{s.t.} \quad & \mathbf{a}(t) = \boldsymbol{\pi}_{\theta}(\mathbf{s}(t))\,,\\
& c(\mathbf{s}(t), \mathbf{a}(t)) \in \mathcal{S},\\
\label{eq:dynamics}
& \boldsymbol{s}(t+1)=f\left(\boldsymbol{s}(t), \boldsymbol{a}(t), \boldsymbol{v}_{e x t}(t)\right);
\end{align}
\end{subequations}
where $\boldsymbol{\pi}_{\theta}(\mathbf{s}_t)$ are the parameters of the voltage control policy, and $\mathcal{S}$ represent the set of safety constraint. $\gamma$ is a discount factor. To be more specific, $c(\mathbf{s}(t), \mathbf{a}(t))$ 
refers to the state safety constraint $\underline{v}_{j} \leq v_{j}(\boldsymbol{p,q}) \leq \bar{v}_{j}, j \in \mathcal{N}$. The underlying dynamics including external variable $\mathbf{v}_{ext}$ is modeled as \eqref{eq:dynamics}. During the training phase, RL agent interacts with the environment driven by the power flow equations, and learns the reactive power dispatch $\mathbf{a} = \pi_{\theta}(\mathbf{s})$ to maximize the accumulated discounted reward. Recent developments of deep RL models have enabled a set of deep learning based algorithms to efficiently solve such reward maximization problem, such as Deep Q Learning for discrete actions (e.g., tap position)~\cite{mnih2013playing}, and Deep Deterministic Policy Gradient (DDPG) for continuous actions~\cite{lillicrap2015continuous}.

However, one intrinsic challenge for DRL policies is to validate the policies given by deep neural networks can always guarantee the voltage staying within the safe region $[\underline{v}_i, \bar{v}_i], i\in \mathcal{N}$. This is due to the fact that during the DRL training stage, the RL agent only focused on maximizing the reward while there is no hard constraint on $\boldsymbol{\pi}_\theta$ to ensure safety. In addition, even if the learned policy appears ``safe'' on the training data (with finite training episodes), it is not guaranteed to be safe under all scenarios. The lack of formal safety guarantees is a major obstacle in the deployment of RL to real-world power systems, as a violation of safe operation constraints can lead to severe impacts such as voltage collapse and cascading failure.
Motivated by the challenge above, our goal is to ensure the RL algorithm obtain provably safety guarantee \emph{during both policy training and policy deployment (after training)}.




\section{Safety Voltage Layer}
\label{sec:simulation}
In this section, we will first discuss how to explicitly obtain the safety constraints for the RL controller with a tractable form. We will then describe our proposed safe controller design. 

To ensure safety during both training and execution, we first need to identify unsafe actions, and then modify the reactive power injections so that the voltage is within safe bounds.  
We achieve this by incorporating the power flow relationship in a compact way for the learned RL agents. In the original DistFlow formulation, we can further neglect the line loss via setting $l_{ij}=0$ for all $(i,j)\in E$. For each node, by assuming $v_{i}\approx 1$, while by approximating $v_j^2-v_i^2\approx 2(v_j-v_i)$, we can get the linearized version of DistFlow model,
\begin{equation}
\label{equ:lin_distflow}
    \begin{gathered}
    -p_{j}=p_{i j}-\sum_{k:(j, k) \in E} p_{j k} \\
    \quad-q_{j}=q_{i j}-\sum_{k:(j, k) \in E} q_{j k}\\
v_{i}-v_{j}= 2(r_{i j} p_{i j}+x_{i j} q_{i j}).
\end{gathered}
\end{equation}

By collecting $\mathbf{v}=[v_1,...,v_n]^T$ and substituting  $p_{ij}$ and $q_ij$ into the last equation of \eqref{equ:lin_distflow}, we can represent the voltage profile in a more compact form
\begin{equation}
\label{eq:linear_approx}
    \mathbf{v}=\mathbf{v}_{0}+\mathbf{R}\mathbf{p}+\mathbf{X}\mathbf{q},
\end{equation}
where $v_0$ is the voltage for the feeder head, and $\mathbf{R}$ and $\mathbf{X}$ are positive matrices.

Once we get the policy $\boldsymbol{\pi}_\theta$ and the safety constraint model for the voltage profile, we want to find the reactive power injections that avoid unsafe voltage deviations. At each control time step, it leads to the following optimization problem
\begin{equation}\label{eq:safety_layer}
    \begin{aligned}
\min _{\boldsymbol{q}} \quad & \frac{1}{2}||\boldsymbol{q}-\boldsymbol{\pi}_\theta (\mathbf{s}(t))||^2_2 \\
\text { s.t. } \; & \boldsymbol{Rp}+\boldsymbol{Xq} + \mathbf{v}_0\leq \bar{\boldsymbol{v}} \\
& \boldsymbol{Rp}+\boldsymbol{Xq} + \mathbf{v}_0 \geq \underline{\boldsymbol{v}}.
\end{aligned}
\end{equation}
By solving \eqref{eq:safety_layer}, we find $\mathbf{q}$ which are not only close to the action given by the deep RL agent, but also satisfy the hard constraints over nodal voltage.

Since \eqref{eq:safety_layer} is a convex, quadratic optimization problem, and we can use standard QP-solvers to solve it efficiently in polynomial time~\cite{ye1989extension}. Alternatively, we can first learn the active constraints, and then use the closed-form solution for the equality constrained problem. If there is at most one constraint that is tight, \cite{dalal2018safe} developed a closed-form solution. However, for voltage control, there is usually more than one constraint on voltage magnitude that can be active. Therefore, we need to solve~\eqref{eq:safety_layer} rather than using the closed-form solution, unless all active constraints can be identified beforehand~\cite{chen2020learning}. In the following, we also give two remarks regarding the linear power flow model and the model knowledge.



\textbf{Remark 1} \emph{Effects of Power Flow Linear Approximation \eqref{eq:linear_approx}:} We note that in our safety layer design, the use of linearized power flow is restricted to validate if the policy $\boldsymbol{\pi}_\theta$ gives safe voltage. Thus we are not losing the representation capability of deep RL on the complex relationship between power injections and voltage. 
Moreover, the linear DistFlow model has been proved accurate verified by real-world data when estimating the voltage box constraints in \eqref{eq:safety_layer}~\cite{lin2021data}. 

\textbf{Remark 2} \emph{Knowledge of Distribution Grid:} Another assumption we make for the proposed safety layer is that we can explicitly write out the inequality constraints involving grid parameters $\mathbf{R}$ and $\mathbf{X}$. Such assumption can be justified when the RL agent has access to historical grid operational data, and can therefore employ machine learning or statistical methods to estimate these parameters.
There are also emerging techniques for estimating both the network topology and line parameters for distribution grid with observational data~\cite{park2018exact,deka2017structure}. We leave the discussion of incorporating model learning error to tighten voltage safety range $[\underline{v}, \overline{v}]$ to future work.

\begin{figure}[t]
	\centering
	\includegraphics[width=0.6\linewidth]{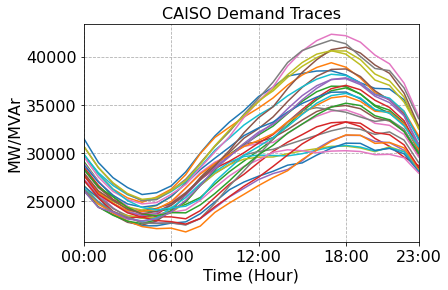}
	\caption{\footnotesize CAISO daily demand data samples used for RL training.}
	\label{fig:caiso_demand}
\end{figure}

\section{Case Study}
\label{sec:simulation}
In this section, we demonstrate the effectiveness of the proposed safety layer method for voltage control. Specially, we show that by incorporating the safety layer in a standard deep deterministic policy gradient (DDPG)~\cite{lillicrap2015continuous} algorithm, we can always find actions satisfying voltage constraints. We compare the proposed safety layer with baseline linear policy or deep RL methods, and validate the improved performance in terms of control costs and safety measures.

\begin{figure*}[]
	\centering
	\includegraphics[width=1.0\linewidth]{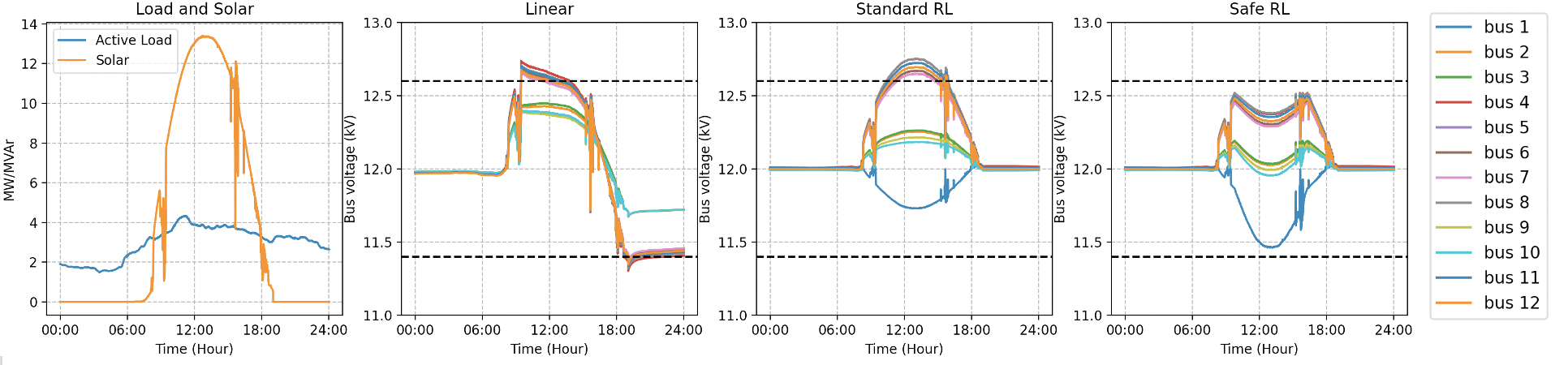}
	\caption{\footnotesize Results on voltage regulation. Dark dashed lines denote the $[\underline{v}_i, \bar{v_i}]$ region. By using proposed safety layer, the resulting reactive power injections enforce the voltage magnitude to be within the safe regions.}
	\label{fig:voltage}
\end{figure*}

\textbf{Experiment Setup} Throughout the simulation, we use the IEEE 13-bus test feeder to validate the algorithm performance. We use real-world load data from CAISO\footnote{\url{https://www.caiso.com/Documents/HistoricalEMSHourlyLoadDataAvailable.html}}, and normalize the demand value based on the 13-bus system configurations. The nominal voltage magnitude at each bus is $12kV$, and we allow $\pm5\%$ voltage deviation for each node. We visualize 20 sample days of training data in Fig. \ref{fig:caiso_demand}.

We employ DDPG algorithm to train the deep RL agent, and we employ 3-layer neural networks for both the policy network and value network in the DDPG agent. For each episode, we collect rollouts for $24$ hours, and we also keep a replay buffer for states throughout RL training process. 


For model comparison, we also use a linear policy to output the reactive power injections. Such linear policy actively considers the voltage magnitude bound, but may take non-optimal control actions which incur larger system costs or short-term voltage violations. We refer readers to \cite{li2014real} for the details of such an algorithm. For the implementation of all the algorithms, we also constrain the reactive power injections within box constraints considering the regulating capabilities of the inverter of each DER. Once we get the control actions from each controller, we use the full power flow model to calculate the resulting voltage. We test the performance of the proposed safe RL approach, standard RL approach, and baseline linear policy using a high-resolution real-world load and renewable data in Fig 3 (left).
Based on the time resolution of the PV and load trajectory, controllers adjust their control output every 6s. We use Pytorch to build all RL models and run the training process. We report the computation time in MacBook Pro Personal Laptop with 16 GB 2400 MHz DDR4 memory and 2.2 GHz Intel Core i7 processor.

\textbf{Simulation Results}
We first compare the voltage profile using three different control schemes. In Fig. \ref{fig:voltage}, we show the nodal voltage magnitude for one day's test data. It can be seen that all three algorithms take actions to try to stabilize the voltage within the safe region ($[11.4kV, 12.6kV]$). But during the middle of the day, both linear controller and standard RL agent lead to greater voltage deviation than the operational limits. The linear policy is relatively ``slow'' in the sense of acting to load and renewables generation change, causing the spikes in voltage profiles. The RL agent finds the control actions using least control efforts (which will be explained in detail in Table \ref{table:result}), but more than half of nodal voltage at noon are exceeding the upper limits. This shows that even though the trained RL agent is able to find control actions to maximize the reward, such training scheme can not exclude unsafe voltage deviations during test time. On the contrary, the proposed safe layer helps maintain voltage staying within the safety bounds. We can further observe that the resulting safe policy manages to reduce the voltages for multiple buses at the same time, meaning that it is possible to refine a trained RL agent to explicitly handle the hard, safety constraints. 

\begin{figure}[t]
	\centering
	\includegraphics[width=0.6\linewidth]{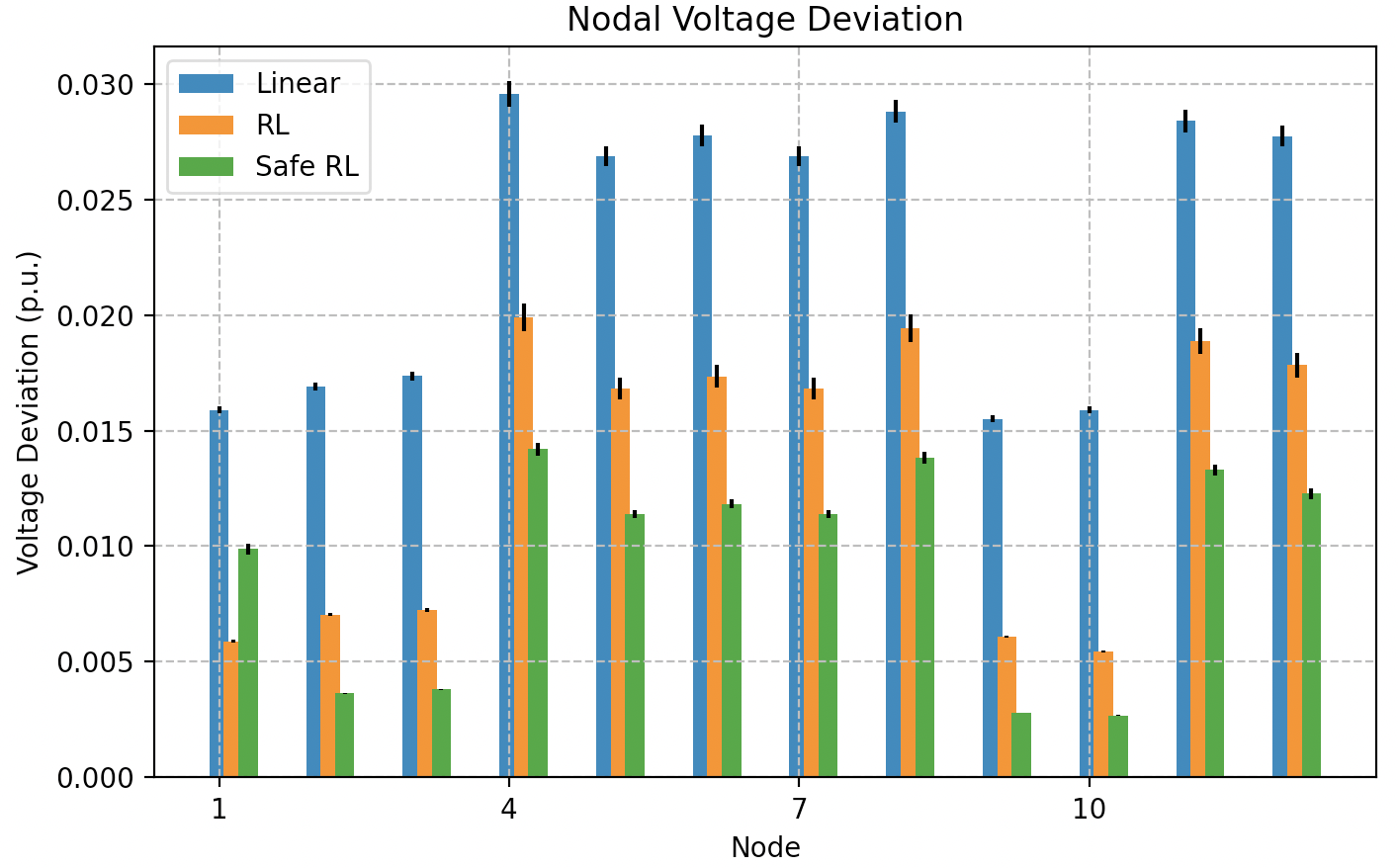}
	\caption{\footnotesize The mean and variance of voltage deviation on each bus under different control schemes.}
	\label{fig:volt_deviation}
\end{figure}

In Fig. \ref{fig:volt_deviation}, we look into the nodal voltage profile, where mean and variance of voltage deviation are plotted for the three methods. We can observe that linear policy leads to voltage profile with the largest deviation across all buses, which shows that the linear model may not achieve satisfactory performance faced with renewables integration in the distribution grid. On average, the safe RL policy can reduce the voltage deviation by more than $30\%$ compared to RL counterparts, showing the necessity of incorporating safety constraints into model-free algorithm design.

\begin{table}[]
\centering
	\renewcommand{\arraystretch}{1.4}
	\begin{tabular}{>{\centering\arraybackslash}m{10em}|>{\centering\arraybackslash}m{9em}|>{\centering\arraybackslash}m{10em}|>{\centering\arraybackslash}m{10em}>{\centering\arraybackslash}m{10em}}
						\ChangeRT{0.7pt}
		Method& Time (s) & Average $q_i$ (kVAR) & $v_i>\%5$ limit\\
				\ChangeRT{0.7pt}
		Linear & $9.02\times 10^{-3}$&1.968 & $9.96\%$  \\
    	RL & $8.88\times 10^{-3}$ &  1.543& $10.82\%$   \\
    	Safe RL &$1.92\times 10^{-2}$ &1.829  & $0.01\%$\\
		\ChangeRT{0.7pt}
	\end{tabular}
	\caption{Statistics for average computation time (per instance), average reactive power injection, and the frequency of infeasible voltages for linear policy, standard RL policy and safe RL policy.}
	\label{table:result}
\end{table}

We report the statistics for solution time, average control efforts, and control results in Table \ref{table:result}. Evaluations are based on simulation results for all $14420$ test data samples. The linear policy and RL agent take nearly the same time to compute the reactive power injections. 
The safe RL does not add much burden to the fast RL policy inference process. The average computation time for the safe RL method is still much smaller than the control step resolution (6s), making it realistic to implement for real-time voltage regulation. This shows that the benefits of the proposed algorithm: with minimal added computational costs, we can find safe policies as a plug-in-play module for off-the-shelf RL algorithms. And all three methods take much shorter time than solving a model-based optimization problem, e.g., by taking SOCP relaxation on the voltage regulation task\cite{chen2020input}. Note that in practice, we can also resort to the safety layer computation to mini-batches for parallel computation, which can further accelerate the solution process of the proposed algorithm.

The effects of the safety layer are further reflected in the average reactive power injection, and the occurrences of unsafe voltage deviation. As is shown in Table \ref{table:result}, safe RL may take more reactive power resources to achieve the regulation task. But compared to both linear policy and standard DDPG agent, the safe RL agent can guarantee for almost all test instances, the resulting voltages are safe. Only for $0.01\%$ of the test samples, safe RL can not find a safe policy, which may be caused by the representation limitation of the linearized power flow model, or the inability to find feasible control actions by only controlling reactive power injections at DER buses.

\section{Conclusion and Discussion}
In this work, we present SAVER, a computationally efficient safe layer for model-free learning-based voltage regulators. By explicitly taking the constraints on voltage magnitude into controller design, we show the resulting controller can always output reactive power injections that guarantee voltage safety. Simulation results demonstrate such safety layer involving linear constraints over the voltage magnitude can be an effective plug-in module for off-the-shelf deep RL algorithms. For future work, we will explore how to autonomously identify the nonlinear relationship between reactive power injections and nodal voltages and ensure safe control at the same time. More advanced safe control schemes based on both active and reactive power injections will also be investigated.

\bibliographystyle{IEEEtran}
\bibliography{references}

\end{document}